\documentstyle[aps,prl,twocolumn,epsf]{revtex}
\begin{document}
\preprint{NHMFL-97}
\title{New Paired-Wavefunction for the Frustrated
Antiferromagnetic Spin-Half Chain}
\author{Z. N. C. Ha}
\address{
National High Magnetic Field Laboratory, Florida State University,
Tallahassee, FL 32306. }
\date{November 24, 1997}
\maketitle
\begin{abstract}
I propose a new paired-wavefunction with a parameter that
{\it continuously} interpolates from the 1D Jastrow-product 
to the Majumdar-Ghosh dimer-wavefunction appropriate for the frustrated 
Heisenberg $S = 1/2$ antiferromagnet.  This spin paired-state
constructed in $S_z$ basis is an alternative 
to the well-known resonating-valence-bond basis state for 
describing the $S = 0$ ground-state with no apparent 
long-range spin order.  Some numerical evidences are presented.
\end{abstract}



\narrowtext

The simplest non-trivial single-particle wavefunction in the first
quantized form is the plane wave, $\exp(i\vec{k}\cdot\vec{r})$,
corresponding to a free particle in space.  For the many-body
system the simplest ground-state wavefunction can be constructed by 
either anti-symmetrizing or symmetrizing the single-particle plane 
waves depending on the statistics of the particles being considered.

The next non-trivial step is to in-cooperate the effects of
interaction by taking (anti-)symmetric pairwise wavefunction and
further (anti-)symmetrizing over all the other remaining variables 
in the many-body wavefunction. The simplest such wavefunctions are the
Jastrow-product \cite{jastrow}
and the less well-known BCS paired-wavefunction projected
onto the Hilbert space of definite number of particles \cite{bcs}. Recently,
these two-types of wavefunctions have been applied to
the fractional quantum Hall states \cite{qhe}. In this letter
I propose a new paired-wavefunction with a parameter
that interpolates from the
Jastrow-product which is an exact ground-state of the Haldane-Shastry
model (HSM) \cite{halsha} to the Majumdar-Ghosh wavefunction for the
frustrated $SU(2)$ Heisenberg spin chain \cite{mg}.

One of the main results of this letter
is the following wavefunction for the
spin chain with $N_a$ sites
\begin{equation}
|\Psi\rangle^l_\lambda = \sum_{\sigma}
\phi^l_\lambda(z,\sigma) e^{i\pi\sum_{j=1}^{M}
m_j}|\sigma\rangle,
\label{wf}  
\end{equation}
where $M = N_a/2$ is the
total number of down or up spins, 
the sum is over all the spin configuration $\sigma$ in $S_z$ basis, 
and $m_i$ ($n_i$) the integer site index of $i$th down(up) spin.
The second factor in the sum corresponds to the Marshall sign.
The orbital function $\phi(z,\sigma)$ is given by
\begin{eqnarray}
\phi^l_\lambda(z,\sigma) &=& \mbox{Det} 
[f^l_\lambda(z^\uparrow_{n_i},z^\downarrow_{m_j})], \\
f^l_\lambda(x,y) & = & (x + y)^{l/M}|x - y|^{\lambda},
\label{pair}
\end{eqnarray}
where $z_{m} = \exp(i2\pi m/N_a)$ and $l$ some integer. 
Here, Det denotes determinant of the $M\times M$ matrix 
$[f(x_i,y_j)]$ while
$f(x_i,y_j)$ is a wavefunction for a pair of up-
and down-spins.  One can consider
$\lambda > 0$ as a measure of {\it apparent} 
repulsion between the spin pair.  

I summarize some noteworthy features of the wavefunction 
as follow:
\begin{itemize}
\item The wavefunction is a $SU(2)$ spin singlet and an
eigenstate of the lattice translation operator with the
wavenumber $Q = 2\pi (M-l)/N_a$.  Hence, $l = 0$ ($l = M$)
corresponds to $Q = \pi$ ($Q = 0)$.
\item When $\lambda = M-1$ and $l = 0$ for odd $M$ 
($\lambda = M - 2$ and $l = M$ for even $M$) the wavefunction 
{\it exactly} reduces to the following fully Gutzwiller-projected 
free-fermion wavefunction \cite{scott}
\begin{equation}
\phi(z,\sigma) = \prod_{i<j} (z^\uparrow_{n_i} - z^\uparrow_{n_j})
\prod_{i<j} (z^\downarrow_{m_i} - z^\downarrow_{m_j}),
\end{equation}
which is the {\it exact} ground-state of the isotropic $SU(2)$
Haldane-Shastry spin chain. The HSM is known to belong to the same
universality class as the Heisenberg spin chain with the 
nearest neighbor exchange \cite{halsha,bethe}.
\item When $\lambda = 1$
it becomes the {\it exact} Majumdar-Ghosh wavefunction with 
definite crystal momentum $Q = \pi$ (for $l=0$) or $Q = 0$ 
(for $l = M$),
\begin{equation}
|\Psi\rangle = |A\rangle \mp |B\rangle,
\end{equation}
where $|A\rangle = (\uparrow\downarrow - \downarrow\uparrow)\cdots
(\uparrow\downarrow - \downarrow\uparrow)$ and $|B\rangle =
T|A\rangle$. ($T$ is the lattice translation operator.)
\item The wavefunction is in the Dyson's form of the
BCS-wavefunction projected onto the Hilbert space of definite
particle number. One major difference here is that the power
$\lambda$ is positive, reflecting {\it apparent} repulsive interaction
between the spin pair.
\end{itemize}

One of the most important features of the proposed wavefunction
mentioned above is the spin singlet property which is also a general
requirement for the ground-state of antiferromagnetically 
correlated spin systems \cite{lieb}.
The wavefunction in Eq.~(\ref{wf}) is by construction $S_z = 0$ state
(i.e. $M = N_a/2$); thus, if $S^+ |\Psi\rangle = 0$ it is also a
$S = 0$ state.
This feature can be shown by proving that
\begin{equation}
|\Psi\rangle + \delta_{\sigma_i,\uparrow} \sum_{j=1}^{N_a} 
\delta_{\sigma_j,\downarrow} P^\sigma_{ij}|\Psi\rangle = 0,
\end{equation}
where $P^\sigma$ is the spin exchange operator.  Using the
antisymmetric property and minor expansion 
of the determinant it is straightforward to
show that the singlet requirement reduces to the following 
condition for some fixed location $z^\uparrow_i$,
\begin{eqnarray}
\phi(z,\sigma) & = &\sum_{j=1}^M (-1)^{i+j} 
f(z^\downarrow_j,z^\uparrow_i){\cal M}^j_i\!\! +\!\!\!
\sum_{j\ne \mu}\!\sum_{\nu(\ne i)} (-1)^{i+j+\mu+\nu} \nonumber \\
&\times&\mbox{sgn}(j-\mu)\mbox{sgn}(\nu-i)f(z^\downarrow_j,z^\downarrow_\mu)
f(z^\uparrow_\nu,z^\uparrow_i){\cal M}^{j\mu}_{i\nu},
\label{cond}
\end{eqnarray}
where ${\cal M}^{i_1i_2\ldots i_p}_{j_1j_2\ldots j_p}$ is 
a $(M-p) \times (M-p)$ matrix obtained from the original $M\times M$ matrix
by eliminating the columns $i_1,i_2,\ldots,i_p$ and the rows 
$j_1,j_2,\ldots,j_p$.  This condition holds true for 
any even function $f$ since the second term in
Eq.~(\ref{cond}) vanishes identically when summed over the dummy indices
$j$ and $\mu$ and the first term becomes just a simple minor expansion of the
original determinant.  Thus, the wavefunction
in Eq.~(\ref{wf}) with $f$ given by Eq.~(\ref{pair})
is a particular case of $SU(2)$ singlet.  It is interesting to note
that the singlet constraint is 
naturally resolved with the BCS-like form of wavefunction in the $S_z$
basis, an alternative to the resonating-valence-bond basis.

The pair function $f$ with $\lambda > 0$ in the wavefunction 
suggests that the interaction between up- and down-spins is of repulsive 
nature.  But, it is somewhat
misleading since for the lattice system there could be an arbitrary
constant factor $\prod_{i<j} (z_i - z_j)^m$ up to a correction to the
phase factor.  Thus, even for the apparently repulsive case of 
$\lambda = M-1$ (or $\lambda = M - 2$), the wavefunction can be divided
by the constant factor with $m=1$ and be rewritten as
\begin{equation}
\phi(z,\sigma) \propto \prod_{ij} 
{\mbox{sgn}(n_i-m_j)\over z^\uparrow_{n_i} - z^\downarrow_{m_j}},
\end{equation}
where the {\it effective} attraction between up- and down-spins 
(i.e. the short-range antiferromagnetic correlation) is made explicit.
It is remarkable that for the case of $\lambda = 1$ 
there also exists an equivalent alternative form given by
\begin{equation}
|\tilde{\Psi}\rangle^l_1 \propto {\cal N} \lim_{\kappa \rightarrow -\infty} 
|\Psi\rangle^l_\kappa,
\end{equation}
where $\cal{N}$ denotes proper normalization.  In this representation the
attraction for a spin pair appears to be infinite.
While the first state corresponds to the ground-state of the
gapless spin liquid where the spinons,
the elementary excitations, are unbound and can have arbitrarily low
energy (i.e. the {\it effective} attraction in this case
is not enough to subdue the pair-breaking quantum fluctuation and to induce
a gap), the second corresponds to the fully paired or dimerized state with gap 
in the energy spectrum.  
I can, thus, infer from the two known cases that as $\lambda$ decreases 
the {\it effective} attraction increases
and that the Majumdar-Ghosh limit ($\lambda = 1$) can be viewed as the 
strongest-possible-pair condensation.  In fact the first case can 
be viewed as the critical point of the frustrated Heisenberg
antiferromagnet given in Eq.~(\ref{ham}) \cite{haldane,fixedpt}, 
and any smaller $\lambda$ corresponds to the massive dimerized phase. 
This transition can be thought of as due to the singlet
pair condensation and that the size of the singlet
pair fluctuation becomes smaller and smaller as the frustration is
increased towards the Majumdar-Ghosh limit.  

In order to show more clearly the nature of pairing 
in the frustrated Heisenberg antiferromagnet 
I start with the following spin Hamiltonian
\begin{equation}
H = \sum_{n=1}^{N_a} \vec{S}_n\cdot\vec{S}_{n+1} + \gamma
\vec{S}_n\cdot\vec{S}_{n+2},
\label{ham}
\end{equation}
where $S = 1/2$ and $\gamma > 0$ is the frustration parameter. The
Hamiltonian with $\gamma = 0$ corresponds to the Bethe-ansatz
solvable model \cite{bethe}.  At $\gamma = 1/2$ the exact doubly
degenerate ground-states are the Majumdar-Ghosh wavefunction\cite{mg}.
When $\gamma$ is greater than a critical value ($\approx 0.241$)
the otherwise marginally irrelevant umklapp processes become relevant
perturbation to the gapless spin-liquid state and the spin-chain
spontaneously dimerizes; and, the critical point is identified with
the $\beta^2\rightarrow 8\pi$ limit of the sine-Gordon model
\cite{haldane}.
Employing the standard method \cite{lp,haldane} one can reduce the
Hamiltonian to the exactly solvable Luttinger
model plus the following umklapp terms expressed with the
Jordan-Wigner lattice fermions,
\begin{eqnarray}
H_u & = & \sum_n\!\!\! \sum_{\delta=1,2}\!\!\!
a_\delta\psi^\dagger_R(n)\psi_L(n+\delta)
\psi^\dagger_R(n+\delta)\psi_L(n) \nonumber \\
&+& \gamma\sum_n \!\!\!\sum_{\delta=\pm
1}\!\!\!\psi^\dagger_R(n)\psi_L(n+\delta)
\psi^\dagger_R(n+\delta)\psi_L(n+2\delta) \nonumber \\
&+& \mbox{Hermitian conjugate},
\label{umklapp}
\end{eqnarray}
where $a_1 = 1$ and $a_2 = -\gamma$.
The rapidly oscillating modes at the Fermi points have already been
factored out from the fermion operators.
When the z-axis anisotropy is allowed 
this umklapp term is also responsible for the N\'eel ordered state
\cite{dennijs}.

Recently, Wilczek and Nayak interpreted 
the umklapp term in the two-dimensional Mott-system
as pairing interaction of particle 
and hole and when the order parameter 
$\langle \psi_L \psi_R^\dagger \rangle$ is a pure imaginary number
a BCS-like state rather than 
the charge-density-wave state is realized\cite{wilczek}. 
In the spin language their BCS-like state corresponds to
the spontaneously dimerised state \cite{haldane}, the
charge-density-wave state to the N\'eel ordered state
\cite{dennijs}, and their particle-hole pairing 
to the spin singlet pairing.  Furthermore, their
BCS-like state through repulsive channel seems 
consistent with our proposed wavefunction with
$\lambda > 0$ which exhibits the
{\it apparent} repulsive interaction.

The original estimate of the critical frustration $\gamma_c = 1/6$
\cite{haldane} which can be obtained from Eq.~(\ref{umklapp}) by 
Taylor expansion of the operators like $\psi(x+\delta)$ is modified 
to $\gamma_c = 0.277$ if the full quantum operator-product expansion 
is used. (Numerical estimate of $\gamma_c$ is approximately
$0.241$ \cite{estimate}.) 
For estimating $\gamma_c$
it is more convenient to change the
dynamical variables to the following canonically conjugate phase
fields
\begin{eqnarray}
\theta(x) &=& 2\pi \int^x dx'(\rho_R(x')+\rho_L(x')), \\
\phi(x) &=& \pi\int^x dx'(\rho_R(x') - \rho_L(x')),
\end{eqnarray}
where $\rho_{R,L}$ are right- and left-density operators
and $[\phi(x),\theta(x')] = i\pi\mbox{sgn}(x-x')$. The original
spin $SU(2)$ operators are represented in these fields as $S^\pm(x)
\propto \exp(\pm i \phi(x))$ and $S_z(x) \propto \exp(\pm i \theta(x))$.
Furthermore, the lattice Jordan-Wigner fermions are decomposed into
the right- and left-moving Mandelstam fermions represented in terms
of the phase fields as follow
\begin{eqnarray}
\psi_R(x) &=& {1\over \sqrt{L}}:e^{i\alpha\phi(x)}::e^{i\beta\theta(x)}:, \\
\psi_L(x) &=& {1\over \sqrt{L}}:e^{i\alpha\phi(x)}::e^{-i\beta\theta(x)}:,
\end{eqnarray}
where $::$ means normal ordering and $\alpha\beta = 1/2$ for the
fermionic statistics. In order to obtain the correct 
anticommutation relations
between the right- and left-moving operators the zero-modes of the
phase fields are separated as $\theta(x) = \theta_0(x) +
\tilde{\theta}(x)$ and $\phi(x) = \phi_0(x) + \tilde{\phi}(x)$ where
$\theta_0(x) = \theta_0 + 2\pi N/L$, $\phi_0(x) = \phi_0 + 2\pi
J/L$ and $[N,\phi_0] = [J,\theta_0] = i$ \cite{haldane2}.  The
operators $\theta_0$, $\phi_0$, $N$, and $J$ are associated with
the bulk modes of the 1D fluid and commute with the $q\ne 0$
bosonic modes. ($N$ here is not the number of particles but rather
the fluctuations from the mean value $N_0$.)  Some care in ordering
the operators is needed when contracting the right- and left-moving
operators and they are given as follow
\begin{eqnarray}
\lim_{(x-y)\rightarrow m a_0} \psi^\dagger_R(x)\psi_L(y) &=& {i\over L}
\mbox{sgn}(x-y)\left(m^2 + \lambda^2\right)^{\tau/2} \nonumber \\
&\times&\left({2\pi a_0\over L}\right)^\tau
:e^{-i2\beta\theta(x)}:,
\label{rl}
\end{eqnarray}
where $\tau = {2\tilde{\beta}^2-\tilde{\alpha}^2/2}$,
$\tilde{\beta} = \beta \exp(\varphi)$, $\tilde{\alpha}
= \alpha \exp(-\varphi)$ and $a_0$ is the lattice constant.
A similar form appears in Ref. \cite{fradkin} but is missing the 
important factor $\mbox{sgn}(x-y)$ and the corrections due to the
short-distance cut-off.
The dimensionless parameter $\lambda$ is introduced to control the
short-distance behavior for normal ordering two exponential
operators and, in this linear
approximation, is fixed by the filling fraction and is given as
$\lambda = 2/(e^C\pi)\approx 0.357$ ($C$ = the Euler's constant).
If $|m| \gg \lambda $ then the control parameter can be dropped
from Eq.~(\ref{rl}), but if the difference is of the order of the
lattice constant it cannot be ignored.

In order to reproduce the correct anticommutation relations at the
non-interacting value of the renormalization constant
$\exp(-2\varphi) = 1$,  I have to set  $\beta = 1/2$, $\alpha
= 1$ and, thus, $\tau = 0$.  More generally, I obtain the following
anticommutation relations for the left and right Mandelstam modes
\begin{equation}
\{\psi_{R}(x),\psi^\dagger_{R}(y)\} = 
\{\psi_{L}(x),\psi^\dagger_{L}(y)\} = Z\delta(x-y),
\end{equation}
where $Z = (\Gamma(\nu -
1/2)/\sqrt{\pi}\Gamma(\nu))\lambda^{2-2\nu}$ with $\nu =
(\tilde{\alpha}/2 + \tilde{\beta})^2$.  Thus, for the free fermions
(i.e. $\exp(\pm \varphi) = 1$) $Z = 1$ or $\nu = 1$. And, this
requirement along with the fermion constraint $\alpha\beta = 1/2$
fix the values of $\alpha$ and $\beta$. In the gapless phase of the
spin chain, however, $\exp(-2\varphi) = 2$ \cite{lp,haldane}. Thus,
$\tilde{\beta}^2
= 1/8$, $\tilde{\alpha}^2 = 2$ and $\tau = -3/4$.
The umklapp terms $H_u$ given in Eq.~(\ref{umklapp}) can now be
contracted to the following form \cite{ha}
\begin{equation}
\left[1-\left(\sqrt{1+\lambda^2\over 4+\lambda^2} +
2{\root 8 \of {4+\lambda^2\over \lambda^2}}\right)\gamma\right]
:\cos(2\theta(x)):,
\label{umklapp1}
\end{equation}
where some overall constants have been dropped. The coefficient
vanishes when $\gamma \approx 0.277$ and changes sign if $\gamma$
further increases. The
difference between this value and the numerical estimate 
is presumably due to the high energy modes with
non-linear dispersion relation near the Brillouin zone boundary.

For half-odd
integer spin chains the Lieb-Schultz-Mattis (LSM) theorem dictates that the 
gap open for both of the two low-energy sectors at the crystal momentum
$0$ and $\pi$ \cite{LSM}.  
And, if the ground-state
is doubly degenerate one can construct two linear combinations that are
not eigenstates of the lattice translation operator.  
Therefore, the dimer gap induced by the singlet
spin-pairing should also accompany the
spontaneously broken translational symmetry. 
In the presence of
holes or in two-dimensional systems, however, 
the spin paired-state could exist without the broken lattice 
translational symmetry.

I now numerically demonstrate that the proposed wavefunction given in 
Eq.~(\ref{wf}) is a good trial wavefunction. I do this 
by taking the overlap of the wavefunction 
with the exact numerically found ground-state of 
$N_a = 16$ spin chain given in Eq.~(\ref{ham}) 
with the frustration $\gamma$ ranging from
$0$ to $1/2$ as shown in Table~\ref{table}.
Listed in the table are the maximum overlap
squared for the sector $Q = 0$ and $Q = \pi$ for four different 
values of $\lambda$ and the corresponding optimum values 
of $\gamma$.
(For $M$ even (odd) and
$\gamma < \gamma_c$ the lowest-energy state in the sector $Q = \pi$
($Q = 0$) is a $S = 1$ state and the $S=0$ state is the next lowest for
any finite-size system. In all other cases of interest
the ground-states for both sectors are spin singlets.)
As shown in Table~\ref{table} the overlap squared is
very close to one although it somewhat deteriorates in the region
$\gamma \approx 0.4$.


In conclusion, I propose a new paired-wavefunction
for the frustrated Heisenberg antiferromagnetic spin chain.
The wavefunction continuously interpolates from the gapless
spin liquid phase to the fully dimerized phase of Majumdar-Ghosh.
From the bosonization method I estimate the critical value of
the frustration and further argue that the gap is induced 
by the singlet pairing in accord with the proposed paired-wavefunction.  
I also give numerical evidences that show the
validity of the paired-wavefunction as ground states of the
frustrated spin chain.  Details will be published elsewhere
\cite{ha}.

I further speculate that an analog of the trial wavefunction given 
in Eq.~(\ref{wf}) adopted to the 2D lattice (i.e. take $z_j$'s as
general complex numbers and use elliptic functions)
is a possible ground-state for the 2D frustrated spin system. 
This $S = 0$ construction
in $S_z$ basis is to be contrasted/compared with the RVB states used,
for example, by Liang et al \cite{liang}.
It is also tempting to conjecture that
the spin-spin singlet pairing correlation survives the doping of the
spin system with small density of holes and that the ``built-in''
superconductivity naturally arises in some reasonable parameter space.
Numerical works are in progress.

The author would like to thank J. R. Schrieffer for inspiration
and general comments and N. Bonesteel and S.C. Zhang for discussions on the
LSM theorem. This work is supported by NSF Grant DMR-9629987 and the
National High Magnetic Field Laboratory.

\begin{table}
\caption{Square of the overlap of the trial wavefunctions given in the
text with the exact $S=0$ ground-states of the spin-chain with $N_a = 16$
sites at the wave-number $Q = 0$ (or $l = M$) and
$Q = \pi$ (or $l = 0$). The frustration $\gamma = 1/2$ corresponds
to the Majumdar-Ghosh point and $\gamma = 0$ to the Bethe-ansatz
solvable Heisenberg spin chain.  The overlaps for
$Q = \pi$ are slightly better than those
for $Q = 0$ in the intermediate region
$0.25 < \gamma < 0.5$. \label{table}}
\begin{tabular}{cccc}
Q & $\lambda$ & $\gamma$ & (Overlap)$^2$ \\
\tableline
0 & 1 & 1/2 & 1 \\
\mbox{}& 3.0 & 0.44 & 0.9795 \\
\mbox{}& 5.0 & 0.29 & 0.9894 \\
\mbox{}& 6.0 & 0.15 & 0.9997 \\
\tableline
$\pi$ & 1 & 1/2 & 1 \\
\mbox{}& 3.0 & 0.41 & 0.9947 \\
\mbox{}& 5.0 & 0.29 & 0.9957 \\
\mbox{}& 7.0 & 0.18 & 0.9994 \\
\end{tabular}
\end{table}

\end{document}